\documentclass[12pt]{article}
\usepackage{simplewick,epsfig,graphics,amsmath,xcolor}

\def\circa#1{\,\raise.3ex\hbox{$#1$\kern-.75em\lower1ex\hbox{$\sim$}}\,}

\newcommand{\Zsl}{Z\hspace{-4.9pt}{\scriptstyle /}}

\newcommand  \f  \varphi

\newcommand{\be}{\begin{equation}}
\newcommand{\ee}{\end{equation}}
\newcommand{\ben}{\begin{displaymath}}
\newcommand{\een}{\end{displaymath}}
\newcommand{\ba}{\begin{eqnarray}}
\newcommand{\ea}{\end{eqnarray}}
\newcommand{\ban}{\begin{eqnarray*}}
\newcommand{\ean}{\end{eqnarray*}}

\begin{document}

\vspace{1.cm}

{\centering

{\Large\bf Lepton Flavor Violation at muon-electron colliders }

\vspace{1.cm}

{\bf \large Fabio Bossi}

{\it INFN - Laboratori Nazionali di Frascati, \\Via Enrico Fermi 54, I-00044, Italy\\
E-mail: fabio.bossi@lnf.infn.it}

{\bf \large Paolo Ciafaloni}

{\it INFN - Sezione di Lecce, \\Via per Arnesano, I-73100 Lecce, Italy \\
E-mail: paolo.ciafaloni@le.infn.it}
\vspace{0.4cm}

}

\vspace{0.3cm}

\begin{abstract}
 Lepton Flavor Violating (LFV) processes are clear signals of physics beyond the Standard Model. We investigate the possibility of measuring this kind of processes at present and foreseeable future muon-electron colliders, taking into account present day bounds from existing experiments. As a model of new physics we consider a Z' boson with a $U'(1)$ gauge symmetry and generic couplings.  Processes that violate lepton flavor by two
units seem to be particularly promising.
\end{abstract}

\vspace{1.cm}

\section{Electron-muon collisions}
Electron-muon collisions have been so far studied using the interaction of a muon beam with a fixed target. 
Muon beams can have energies up to several hundreds GeV, fluxes of 10$^7$-10$^8$ particles per second and
transverse dimensions of order of a few centimeters \cite{BeamSPS,BeamFNAL,BeamJPARC}.
Although with this technique one can in principle reach very high luminosities by the use 
of properly studied targets, the available center of mass energy (c.m.e.) is strongly suppressed because of the Lorenz boost.

In recent years, however, a very intense R$\&$D program has been put forward in many laboratories to develop 
techniques to obtain very high energy muon-muon interactions in collider mode,
 motivated by the possibility
of producing collisions of point-like particles at very high energy without the limitation from synchrotron 
radiation typical of electron-positron machines. 
In these studies muons are produced either as decay products of pions from fixed-target
 proton-proton collisions, or as a result of pair production in electron-positron collisions. 
Typically, at c.m.e. of 6 TeV, two single bunches
of 2$\times$10$^{12} $ muons, each of transverse dimensions of $\sim$1.5 $\mu$m, collide at a rate of $\sim$50 kHz, 
providing a peak luminosity largely exceeding   10$^{34}$ cm$^{-2}$s$^{-1}$ \cite{BoscoloMC,PalmerMC,MC6Tev,Lemma}. 
One of the main technical limitations to obtain high luminosity comes from the difficulty in producing highly collimated muon beams. A big step forward in this direction has been recently reported by the MICE collaboration which has been able to confirm the success of a ionization cooling experiment on a low momentum muon beam using properly 
designed absorbers \cite{Mice}. 
This experiment has to be considered as an important advance in the development of high brightness muon beams. 

In this paper we want to call the attention to  the fact that such beams could also be used to produce
 electron-muon collisions in collider mode, by the simultaneous usage of high energy electron/positron beams.
This would allow probing electron-muon interactions at c.m.e orders of magnitude higher with respect to 
the fixed target option.

Actually, electron and positron beams with energies up to 100 GeV have been already put in collision at LEP \cite{LEP}.  
Projects for the construction of future linear
or circular e$^+$e$^-$ colliders with c.m.e. between 200 and 1500 GeV and luminosities exceeding 
10$^{34}$ cm$^{-2}$s$^{-1}$
have been also put forward, limited mostly by budget issues. 
These projects vary a lot among each other  in the strategy to reach high luminosity, which is obtained 
either by maximising the rate of collisions ($f_{c}$) or the number of particles in each bunch ($N_{l}$)
or by minimising the beams dimensions ($\sigma_{x,y}$); typical figures of merit are 
 $ f_{c} \geq$ 100 kHz, $N_{l} \sim 10^{11}-10^{12}$,$\sigma_{x,y} \leq 1\mu$m \cite{ILCtdr,FCCee,CEPC}.

For  two colliding beams of particles of type A and B the luminosity can be computed to good approximation 
 by the formula:
\be
 \frac{N_A \cdot N_B}{2 \pi \sqrt{\sigma_{Ax}^{2} + \sigma_{Bx}^{2}} \sqrt{\sigma_{Ay}^{2} + \sigma_{By}^{2}}
 } \times f_{c}
\ee

Using the numbers quoted above we see that, by  properly combining  the aforementioned lepton
beams,
it is at least in principle conceavable to obtain electron(positron)-muon collisions at a 
luminosity of 10$^{34}$ cm$^{-2}$s$^{-1}$ or more, in a c.m.e. range between few hundreds to few thousands GeV. 

A thorough discussion of the technical details of  this hypothetical machine goes well beyond the purpose of our paper. We note, however, that since the collision rate would likely be limited by the revolution rate of the muon beam, reaching  high luminosity must rely on the ability of producing very compact and dense lepton bunches.  We note also that  the scheme in \cite{Lemma}  naturally provides a source for both muon and positron beams, although the latter must be accelerated at higher energies in a subsequent stage. 

\section{Physics case}
Although in principle many interesting measurements can be performed with the above mentioned machine, we focus
 our attention on the possibility of studying
e$^{+(-)} \mu^{-(+)} \rightarrow $ e$^{-(+)} \mu^{+(-)}$  transitions. 
The observation of this process would be a clear
signature of Lepton-Flavor-Violation (LFV). In the Standard Model (SM)  
 LFV processes are suppressed to an unobservable level, therefore the observation of the aforementioned events would   be a clear signature of physics 
Beyond the Standard Model (BSM). 

Many specific and well motivated BSM
 models including LFV can be found in literature (see for instance \cite{scarpa1}; see also 
 \cite{scarpa2} and references therein). We are however not 
interested in discussing them in detail, but will rather use a simple model where LFV transitions are 
mediated by a generic heavy neutral boson (Z') with mass and coupling to be determined by the experiment.
Our purpose is, in fact, to give an order-of-magnitude estimate of the potentials and the limits of the
 proposed experiment as a function of the reachable c.m.e. and luminosity.   

We introduce a new LFV interaction mediated by a neutral Z' boson of mass M$_{z'} \gg$  M$_{z}$, described by the
 following lagrangian:

\be\label{eq:lagr}
 g^{eL}_{ij}\bar{e_i}
\Zsl' P_Le_j+ g^{eR}_{ij}\bar{e_i}
\Zsl' P_Re_j+g^{\nu L}_{ij}\bar{\nu_i}
\Zsl' P_L \nu_j+ g^{\nu R}_{ij}\bar{\nu_i}
\Zsl' P_R\nu_j
\ee
       
We take the Z' to be a gauge singlet; then SU(2)$_L$ invariance implies that g$^{eL}_{ij}$=g$^{\nu L}_{ij}$. We do not make any assumption on the couplings of the Z' with quarks, therefore our results are not affected by the recent LHCb indications for the possible violation of lepton universality \cite{Lhcb}. 
We only note that this result can be explained through a  LFV Z', as discussed in \cite{Crivellin:2015era}. 
 
\section{Low energy bounds on Z' couplings}
\begin{figure}[htb]
\begin{center}
\includegraphics[width=12cm]{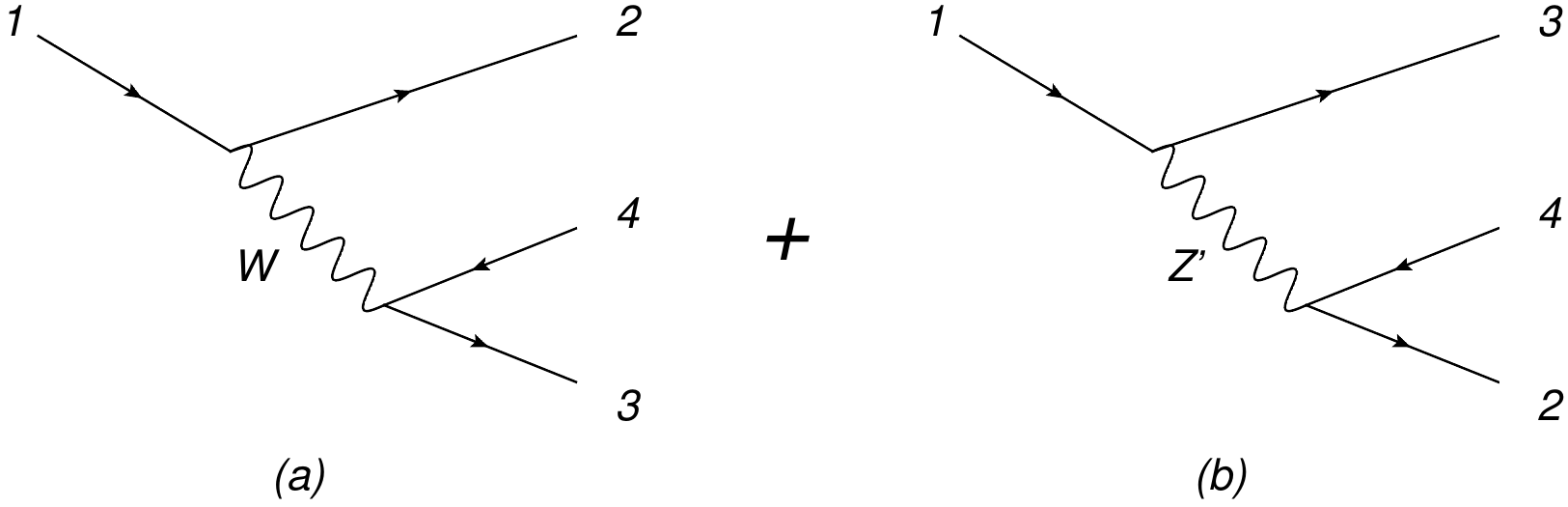}
\caption{\textit{ W and Z' exchange diagrams for the amplitude of a generic decay $1\to 2+3+4$. In the case of `standard' muon decay we have $1=\mu^-,2=\nu_\mu,3=e^-,
4=\bar{\nu}_\mu$. In the case of LFV muon decay, $1=\mu^-,2=e^-,3=e^-,
4=e^+$; in this case only the Z' exchange diagram contributes. }}
\label{fig:gamma}
\end{center}
\end{figure}

Z' exchange amplitudes generate LFV violating processes like 
$\mu^-\to e^-e^-e^+$, but also contribute to `standard' non LFV-violating processes
like muon decay $\mu^-\to e^-\bar{\nu}_e\nu_\mu$ (see fig \ref{fig:gamma}).
In the latter case, the decay width is given by the sum of three terms corresponding to W exchange square, Z' exchange square and interference; we obtain:

\be\label{eeeo}
\frac{\Gamma_\mu }{m_\mu^5}=\frac{G_F^2}{192\pi^3}
-\frac{4\sqrt{2}}{1536\pi^3}\frac{G_F(g^L_{\mu e})^2}{M_{Z'}^2}
+\frac{[(g^L_{\mu e})^2+(g^R_{\mu e})^2][(g^L_{\mu e})^2+(g^R_{\nu_\mu \nu_e})^2]}{1536\pi^3M_{Z'}^4}
\ee

\begin{figure}[htb]
\begin{center}
\includegraphics[width=14cm]{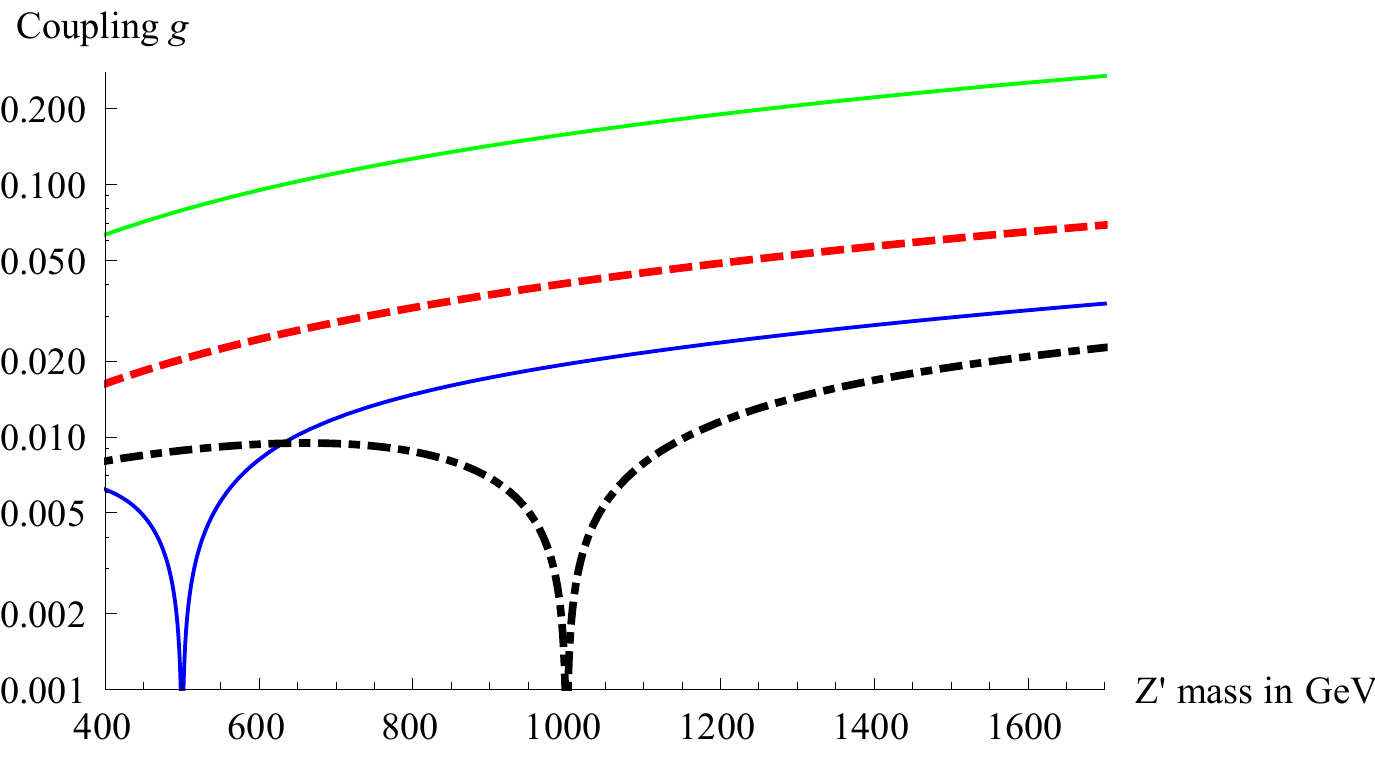}
\caption{\textit{ Upper bound on $g_{\mu e}$ as a function of $M_{Z'}$ coming from muon decay (red dashed curve) and from muonium-antimuonium oscillation (green continuous line). Value of $g_{\mu e}$ that corresponds to one event/year as a   function of $M_{Z'}$  with luminosity ${\cal L}$=10$^{34}$ cm$^{-2}$s$^{-1}$ and  c.m.e. of 500 GeV (blue continuous line) and 1 TeV (dot-dashed line). Left and right couplings are assumed to be equal.} }
\label{fig:main}
\end{center}
\end{figure}

Corrections to the Fermi constant, defined through this process, must be below the per-mille level in order to avoid conflicts with electroweak precision data. We conservatively demand the impact on the Fermi constant to be one order of magnitude smaller\footnote{A more refined analysis would imply a precise evaluation of the impact of the Z' contribution on the electroweak precision tests; this goes beyond the scope of the present paper.}, namely:
\be 
|BR(\mu^-\to e^-\bar{\nu}_e\nu_\mu)-BR(\mu^-\to e^-\bar{\nu}_e\nu_\mu)_{SM}|
\le  \times 10^{-4}
\ee
Notice that one can basically  evade the bounds coming from this request by making suitable assumptions on the chirality of the couplings. Indeed, if the Z' mass is of the order of 1 TeV as we assume in the present paper,  the third term in 
(\ref{eeeo}) is suppressed by $\frac{M_W^2}{M_{Z'}^2}\approx 10^{-2}$ with respect to the second one. If the $\mu$-e-Z' coupling is purely right ($g^L_{\mu e}=0$), then the second term is 0 and we obtain very weak bounds on $g^R_{\mu e}$. Howewer, in order to be conservative, we assume instead all couplings to be the same
$g^{L,R}_{ij}=g\forall i,j$ and obtain stronger bounds. The resulting upper bound on $g$ is plotted on the dashed red line in fig. \ref{fig:main}. Moreover, we have considered the bounds on g coming from muonium-antimuon oscillation.
Using the effective Lagrangian generated by eq (\ref{eq:lagr})  we obtain the oscillation probability
 \cite {Weinberg,Masiero}:
\ben
P(\bar{Mu}-Mu)\approx 10^{-9}\left(\frac{1TeV}{M_{Z'}}\right)^4|4(g^L_{\mu e})^2+4(g^R_{\mu e})^2)|^2
\een
Comparing  with the present experimental bound \cite{Willmann:1998gd}
\ben
P(\bar{Mu}-Mu)\le 8.2 \times 10^{-11}
\een
we obtain:
\be\label{iuoih}
|(g^L_{\mu e})^2+(g^R_{\mu e})^2)|\circa{<}0.07 \left(\frac{M_{Z'}}{1TeV}\right)^2
\ee

This bound is drawn in fig. \ref{fig:main} together with the bound coming from muon decay assuming that left and right couplings are equal. Notice however that the abovementioned bounds depend on different combinations of the couplings (compare (\ref{eeeo}) with (\ref{iuoih})): for instance, if $g_{\mu e}^L=0$ the bound from muon decay is evaded but muonium-antimuniom oscillation still puts bound on the value of $g_{\mu e}^R$.

Let us now discuss the bounds coming from the LFV violating process 
$\mu^-\to e^-e^+e^-$, for which the current experimental bound 
\cite{pdg} is $BR<10^{-12}$. 
In this case only the Z' exchange contributes and assuming $g^L_{\mu e}=g^R_{\mu e}=g_{\mu e}, g^L_{e e}=g^R_{e e}=g_{e e}$ we obtain:
\be\label{cidecuh}
\Gamma(\mu^-\to e^-e^+e^-)=m_\mu^5 \frac{(g_{ee}g_{\mu e})^2}{384\pi^3M_{Z'}^4}
\ee
This process sets therefore bounds on the product of diagonal ($g_{ee}$) and non diagonal ($g_{\mu e}$) couplings. In the following we assume that $g_{ee}$, is sufficiently small, i.e. it respects the bounds coming from $B.R<10^{-12}$ and we focus our attention on the non diagonal couplings. For instance if $g_{\mu e}\approx 10^{-2}$ (see fig \ref{fig:main}), we obtain $g_{ee}<10^{-3}$.


\section{e$^{+} \mu^{-} \rightarrow $ e$^{-} \mu^{+}$  and e$^{-} \mu^{+} \rightarrow $ e$^{+} \mu^{+}$ }
\begin{figure}[htb]
\begin{center}
\includegraphics[width=12cm]{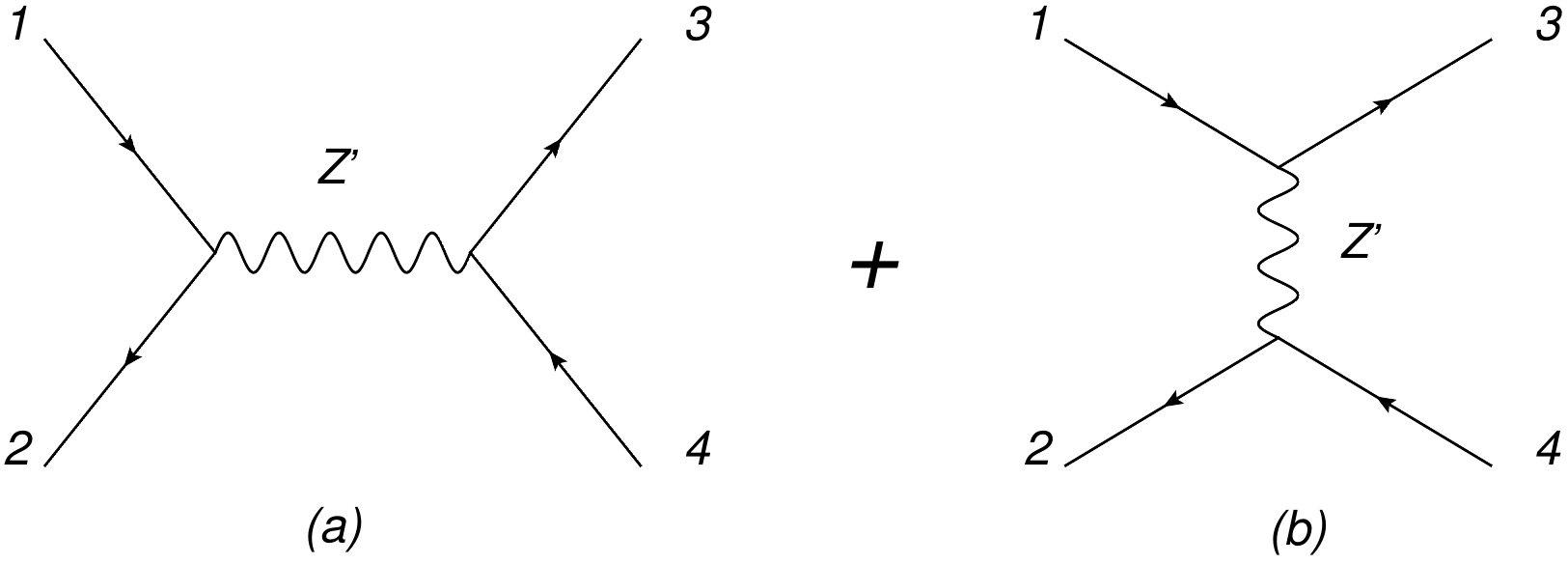}
\caption{\textit{ s- and t- channels contributions to the amplitude for a generic scattering cross sections.}}
\label{fig:sigma}
\end{center}
\end{figure}

The e$^{+(-)} \mu^{-(+)} \rightarrow $ e$^{-(+)} \mu^{+(-)}$ transition violates lepton flavor by two units and  is sensitive only to g$^{eR}_{\mu e}$ and 
g$^{eL}_{\mu e}$. 
The cross section can be computed as the sum of three
different contributions corresponding to the exchange of a Z' in the s channel, in the t channel and to the 
interference between the two, see figure \ref{fig:sigma}. In general, the cross sections for the LFV-violating processes we consider in the present paper are obtained by summing Z' exchange in s- and t- channel (see figure \ref{fig:sigma}). Considering the process $1 2\to 3 4$ and indicating with $g^R_{ij},
g^L_{ij}$ the relevant Z'-fermion-antifermion couplings, the differential cross section in
the scattering angle $\theta=\hat{p}_1\cdot\hat{p}_3$ is the sum of three contributions coming from s-channel square, t-channel squared and s-t interference:
\ba\label{eq:1}
s\frac{d\sigma_{ss}}{d\cos\theta}
&=& \frac{s^2}{8\pi(s-M^2)^2}\left\{
\frac{ t^2}{s^2}[(g^R_{12})^2(g^L_{34})^2+(g^L_{12})^2(g^R_{34})^2]
+    
\frac{u^2}{s^2}[(g^L_{12})^2(g^L_{34})^2+(g^R_{12})^2(g^R_{34})^2]
\right\}
\\\label{eq:2}
s\frac{d\sigma_{tt}}{d\cos\theta}
&=& \frac{s^2}{8\pi (t-M^2)^2}\left\{
 [(g^R_{13})^2(g^L_{24})^2+(g^L_{13})^2(g^R_{24})^2]
+    
\frac{u^2}{s^2}[(g^L_{13})^2(g^L_{24})^2+(g^R_{13})^2(g^R_{24})^2]
\right\}
\\\label{eq:3}
s\frac{d\sigma_{st}}{d\cos\theta}
&=& \frac{s^2}{4\pi(s-M^2)(M^2-t)}\left\{
\frac{ u^2}{s^2}[g^L_{12}g^L_{13}g^L_{24}g^L_{34}+
g^R_{12}g^R_{13}g^R_{24}g^R_{34}]
\right\}
\ea
The expression for the total cross section is rather cumbersome, but it simplifies 
in the case $g^R_{ij}=g^L_{ij}=g\forall i,j$:
\be
\sigma_{TOT}=
\frac{g^4}{3\pi M^2}\frac{6\rho^2(1-\rho)(1+\rho)^2\log\frac{1+\rho}{\rho}
+6\rho^4+3\rho^3-5\rho^2-5\rho+3}{(1-\rho)^2(1+\rho)};\qquad
\rho\equiv\frac{M^2}{s}
\ee
In the case of heavy Z', $\sqrt{s}\ll M_{Z'}$, this expression further simplifies to the following approximate one:
\be
\label{xsecheavy}
\sigma_{TOT}=\frac{2g^4}{3\pi}\frac{s}{M_{Z'}^4}+{\cal O}(\frac{1}{M_{Z'}^6})
\ee

The cross section for the process e$^{+(-)} \mu^{-(+)} \rightarrow $ e$^{-(+)} \mu^{+(-)}$ can be computed
using (\ref{eq:1}-\ref{eq:3}) with  $1=e^-,2=\mu^+,3=\mu^-,4=e^+$. If we further assume  left and right couplings to be equal ($g^L_{\mu e}=g^R_{\mu e}\equiv g$) we can compute the single event sensitivity (s.e.s.) of our proposed experiment, defined as the value of $g$ such to have 1 event/year. As an example
 we plot in fig. \ref{fig:main}  the s.e.s. as a function of the Z' mass at our $\mu-e$ collider, assuming two possible c.m. energy $\sqrt{s}=$ 500 GeV and 1 TeV, and luminosity ${\cal L}=10^{34} cm^{-2} s^{-1}$. In both cases, the explorable region of couplings lies a factor 2-3 below the pessimistic limit set by the standard muon decay\footnote{We call the limit given by the red dashed line in fig \ref{fig:main} `pessimistic' since a more detailed analysys including the possibility of different left and right couplings, a refined estimate of the calculation of the impact on precision electroweak tests would push the red line towards higher values. }. Here, we do not make any assumption on the actual detection efficiency; we note only, however, that it should be taken not below $\sim$50$\%$, with this requirement being less stringent with increasing c.m. energies.  

\begin{figure}[htb]
\begin{center}
\includegraphics[width=12cm]{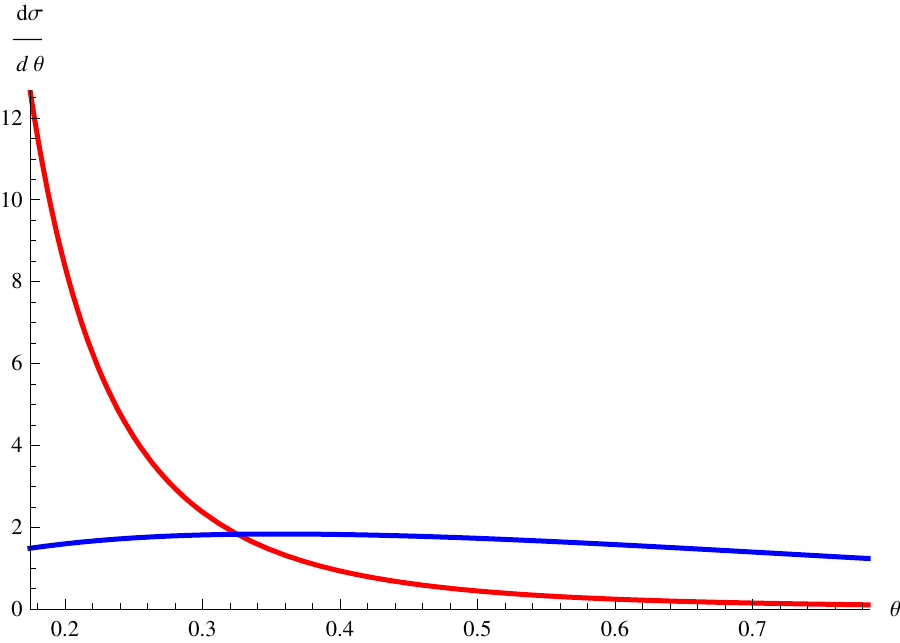}
\caption{\textit{Muon  angular distribution for e$^{+(-)} \mu^{-(+)} \rightarrow $ e$^{-(+)} \mu^{+(-)}$ in the lab frame, where the minitial muon energy is  3 TeVs and the initial electron one is 200 GeVs. The red curve corresponds to a Z' mass of 100 GeV, while the blue one corresponds to a Z' mass of 3 TeVs.}}
\label{distrib}
\end{center}
\end{figure}

The observables discussed until now are not sensitive directly to the Z' mass, but rather to the combination $\frac{g}{M_{Z'}}$ (see (\ref{eeeo},\ref{cidecuh})). We point out 
that the angular distribution of the outgoing (anti)muon is, instead, sensitive to the value of the Z' mass alone. This can be seen from fig. \ref{distrib} where this distribution in the laboratory frame of a colliding muon of energy 3 TeV and an electron of energy 200 GeV is plotted. In order to emphasize the dependence on the Z' mass we take the   values  of $M_{Z'}$=3 TeV for the blue curve and $M_{Z'}$=100 GeV for the red curve.
The reason for the pronounced dependence on the Z' mass is the following. The amplitude is the sum of s-channel contribution, proportional to $1/(M_{Z'}^2-s)$, and t-channel proportional to $1/(M_{Z'}^2-t)$, with $-s\le t\le 0$. If the Z' is heavy, $M_{Z'}^2\gg s$, 
both amplitudes are proportional to 1/$M_{Z'}^2$ and angle-independent, therefore the angular distribution is basically flat (blue curve). Indeed, from (\ref{eq:1},\ref{eq:2},\ref{eq:3}) for couplings all equal to $g$, we obtain:
\ben
\frac{d\sigma_{tot}}{d\cos\theta}=\frac{g^4s}{4\pi M_{Z'}^4}\left(1+\frac{(1-\cos\theta)^2}{4}\right)
\een
which is a rather smooth angular distribution. 
If instead the Z' is light, $M_{Z'}^2\ll s$, then for small angles the distribution is dominated by the t-channel amplitude,
which is very peaked in the forward direction being roughly proportional to $1/t$ 
where $t=-s/2(1-\cos\theta)$ goes to 0 with the scattering angle. In this case we obtain:
\ben
\frac{d\sigma_{tot}}{d\cos\theta}\approx\frac{g^4s}{4\pi M_{Z'}^4}\frac{1}{\left(\frac{1}{2}(1-\cos\theta)+\frac{M_{Z'}^2}{s}\right)^2}
\een
which is instead, for small values of $\frac{M_{Z'}^2}{s}$, strongly peaked at $\theta\approx 0$.
\section{Other processes}
We want now to briefly comment on the possibility to perform two other  experiments.

Firstly, we consider the possibility to observe the LFV process  e$^{+(-)} \mu^{-(+)} \rightarrow $ e$^{+} e^{-}$ at 
the proposed collider. This transition is crossing-connected to the decay $\mu^-\to e^-e^-e^+$, that therefore imposes strong bounds on the couplings. Let us assume $g^{L,R}_{\mu e}=g_{\mu e}$ and $g^{L,R}_{ee}=g_{ee}$; moreover $\sqrt{s}\ll M_{Z'}$. Using 
equation(\ref{cidecuh}) and imposing the experimental limit on the  $\mu^-\to e^-e^-e^+$ branching ratio ($\overline{BR}$), we get:
\be
\frac{g_{ee}^2 g_{\mu e}^2}{2G_F^2 M_{Z'}^4}<\overline{BR},\qquad
\sigma(e^{+(-)} \mu^{-(+)} \rightarrow  e^{+} e^{-})=\frac{2g_{ee}^2 g_{\mu e}^2s}{3\pi M_{Z'}^4}\Rightarrow
\sigma < 
\frac{4 }{3 \pi}  \,\overline{BR}\, G_F^2s
\ee

We see that,  for instance, for $\sqrt{s}$= 500 GeV, $\sigma$\circa{<}0.5$\times$10$^{-8}$ pb, which makes the 
process impossible to be observed for any reasonable value of the collider's luminosity.  Notice that this holds true whatever the value of the couplings: indeed, both the cross sections and the branching ratios depend on the same combination of couplings.

Secondly, we discuss the potential for observing the process  e$^{+(-)} \mu^{-(+)} \rightarrow $ e$^{-(+)} \mu^{+(-)}$ with a fixed target experiment using presently available muon beams. We take as a reference the setup of \cite{MuonE}. Here, 150 GeV muons  impinge on the atomic electrons of a Be target, with total thickness of 60 cm. The beam flux is $\sim$5$\times$10$^{7} \mu$/s. With respect to \cite{MuonE}, we further assume that a magnetic spectrometer allows the discrimination between particles of different charge. Given the numbers quoted above the c.m.e of the process is 400 MeV and the experiment's luminosity $\sim$ 10$^{33}$cm$^{-2}$s$^{-1}$. We must here impose the bounds coming from standard muon decay, similarly to what has been done in the previous section for the high c.m. energy case. Unfortunately  in this case the bounds thus obtained for the values of the couplings are at least one order of magnitude smaller than those required to have one event per year\footnote{A new experiment has also been proposed in \cite{Koike:2010xr} where the process $\mu^- e^-\to e^-e^-$ is enhanced by $(Z-1)^3$, $Z$ being the atomic number. However no additional experimental result can change the conclusions of this section.}. Therefore (maybe unless the luminosity is strongly improved), the LFV processes we consider here are unobservable at present day colliders. 

\section{Conclusions}
In recent years, 
relevant progresses have been made in the design for a future muon collider capable of reaching center of mass energies of several TeV. This machine, which would allow the search of phenomena beyond the Standard Model into a so far unexplored energy domain, requires  producing very intense, high energy muon beams, with collimation characteristics never obtained to date. We suggest that these beams can also be used to produce electron-muon collisions at very high energies, with luminosity exceeding   10$^{34}$ cm$^{-2}$s$^{-1}$. We have shown that such a machine, would allow studying the lepton flavor  violating process    e$^{+-} \mu^{-+} \rightarrow $ e$^{-+} \mu^{+-}$ with a sensitivity potentially better than the one reached by present or proposed rare decay experiments. We have also studied the possibility of performing the same experiment exploiting presently available muon beams, such as that used by the MuonE experiment; we have shown that, unfortunately, in this case the reachable luminosity is at least one order of magnitude smaller than that required to unequivocally observe the  signal of our interest.  Finally, the abovementioned process, that violates LF by 2 units, looks like the most promising one. For instance, processes that violate LF by 1 unit like $e^-\mu^+\to e^-e^+$, are impossible to observe given the bounds coming from present day experiments.

\end{document}